
%
\input phyzzx
\catcode`@=11
%
%
\newtoks\KUNS
\newtoks\HETH
\newtoks\monthyear
\Pubnum={KUNS~\the\KUNS\cr HE(TH)~\the\HETH}
\KUNS={1198}
\HETH={93/03}
\monthyear={May, 1993}
\def\p@bblock{\begingroup \tabskip=\hsize minus \hsize
    \baselineskip=1.5\ht\strutbox \topspace-2\baselineskip
    \halign to\hsize{\strut ##\hfil\tabskip=0pt\crcr
    \the\Pubnum\cr hep-th/9309150\cr \the\monthyear\cr }\endgroup}
\def\bftitlestyle#1{\par\begingroup \titleparagraphs
    \iftwelv@\fourteenpoint\else\twelvepoint\fi
    \noindent {\bf #1}\par\endgroup}
\def\title#1{\vskip\frontpageskip \bftitlestyle{#1} \vskip\headskip}
%
%
\def\acknowledge{\par\penalty-100\medskip \spacecheck\sectionminspace
    \line{\hfil ACKNOWLEDGEMENTS\hfil}\nobreak\vskip\headskip}
%
%

%
\def\journal#1&#2(#3){\begingroup \let\journal=\dummyj@urnal
    \unskip, \sl #1\unskip~\bf\ignorespaces #2\rm
    (\afterassignment\j@ur \count255=#3) \endgroup\ignorespaces}
\def\andjournal#1&#2(#3){\begingroup \let\journal=\dummyj@urnal
    \sl #1\unskip~\bf\ignorespaces #2\rm
    (\afterassignment\j@ur \count255=#3) \endgroup\ignorespaces}
\def\andvol&#1(#2){\begingroup \let\journal=\dummyj@urnal
    \bf\ignorespaces #1\rm
    (\afterassignment\j@ur \count255=#2) \endgroup\ignorespaces}
\def\NP{Nucl.~Phys.}
\def\PL{Phys.~Lett.}
\def\PR{Phys.~Rev.}
\def\PTP{Prog.~Theor.~Phys.}
\catcode`@=12
%

\titlepage

\title{Another Perturbative Expansion
       \break in Nonabelian Gauge Theory}

\author{Izawa K.-I.}

\address{Department of Physics, Kyoto University
         \break Kyoto 606, Japan}

\abstract{
We consider a new perturbation scheme in nonabelian gauge theory.
Pure Yang-Mills theory in three dimensions
is taken as a concrete example.
The zeroth-order in the perturbative expansion
is given by BF theory coupled to a St{\" u}ckelberg-like field.
The effective coupling for the expansion can be small
in the infrared regime, which implies
that nonperturbative treatment of Yang-Mills theory
may be partially reduced to that of BF theory.
}

\endpage


\font\sc=cmr5 scaled\magstep1

\def\a{\alpha}
\def\d{\delta}
\def\e{\epsilon}
\def\k{\kappa}
\def\m{\mu}
\def\n{\nu}
\def\r{\rho}
\def\z{\zeta}
\def\D{\Delta}
\def\L{\Lambda}

\def\p{\partial}

\def\da{\delta_a}
\def\db{\delta_b}
\def\brs{\delta_{\hbox {\sc B}}}


\REF\Bir{For a review, D.~Birmingham, M.~Blau, M.~Rakowski, and
         G.~Thompson \journal \nextline Phys.~Rep. &209 (91) 129.}

\REF\Abe{M.~Abe and N.~Nakanishi \journal \PTP
         &89 (93) 501.}

\REF\Nak{M.~Abe and N.~Nakanishi \journal \PTP
         &85 (91) 391; \andvol &88 (92) 975.}

\REF\Oda{I.~Oda and S.~Yahikozawa, preprint IC/90/44.}

\REF\Hat{H.~Hata \journal \PL &B143 (84) 171.}

\REF\Kug{H.~Hata and T.~Kugo \journal \PR &D32 (85) 938.}

\REF\Oji{T.~Kugo and I.~Ojima \journal
         Suppl.~\PTP &66 (79) 1; \nextline
         N.~Nakanishi and I.~Ojima, {\sl Covariant Operator
         Formalism of Gauge Theories and Quantum Gravity}
         (World Scientific, 1990); \nextline
         See also H.~Hata and I.~Niigata
         \journal \NP &B389 (93) 133.}

\sequentialequations

%
{\caps 1. Introduction}

Perturbation theory has been successful in quantum field theory
when the effective coupling is small in the energy region considered:
two main examples are provided by
electroweak theory in the accessible energy range and
quantum chromodynamics (QCD) in the regime of deep inelastic
scattering.
However, naive perturbation theory cannot be applied
to QCD in the infrared regime,
where the effective coupling is expected to be large.
The conventional perturbative expansion
is based on free field theory, which seems inappropriate
for nonabelian gauge theory in the confining phase.

The question we address in this paper is
whether there exists any perturbation theory
appropriate for QCD with the effective coupling for the expansion
small in the infrared regime.

In the investigation of topological field theory,
it was noticed that BF theory
\refmark{\Bir}
can be regarded as a zero-coupling limit of Yang-Mills (YM) theory,
especially in two dimensions.
In higher dimensions, however, the limit is singular
due to the fact that the gauge symmetry in BF theory
is larger than that in YM theory.

Quite independently, Abe and Nakanishi have recently claimed
\refmark{\Abe}
that BF theory is essentially equivalent to the zeroth-order
approximation to YM theory in their new method
\refmark{\Nak}
of solving gauge theories
in the covariant operator formalism.
Their perturbative expansion is made at the operator level:
namely, field operators themselves are expanded
in powers of the coupling,
and they are to be solved by means of field equations and
equal-time commutation relations.

Inspired by these observations,
we propose another perturbation scheme in nonabelian gauge theory.
For simplicity, in this paper, we deal with pure YM theory in three
spacetime dimensions.
The zeroth-order in the perturbative expansion
is given by BF theory coupled to a St{\" u}ckelberg-like field.
The effective coupling for the expansion can be small
in the infrared regime, which implies that nonperturbative
treatment of YM theory may be partially reduced to that of BF theory.

\endpage

%
{\caps 2. The Model Lagrangian}

In this section, we present BF-theory-like formulation
of pure YM theory in three dimensions.
The purpose of reformulating naive YM Lagrangian
into BF-theory-like one is to remove the singularity
which appears in the zero-coupling limit of YM theory
due to gauge-symmetry enhancement stated in the Introduction.

Let us start from a Lagrangian
$$
 \eqalign{
  {\cal L}_S &= {1 \over 2}\e^{\mu \nu \r}B_\m F_{\n \r}
                 + {1 \over 2}\k^2B_\m^2; \cr
  F_{\m \n} &= \p_\m A_\n -\p_\n A_\m
                - ig[A_\m, A_\n], \cr
 }
 \eqn\START
$$
where the vector fields $A_\m$ and $B_\m$ take values in
a simple Lie algebra $\cal G$,
and $g$ and $\k$ denote non-vanishing coupling constants.
This Lagrangian is equivalent to the usual YM Lagrangian,
as is clear when the field $B_\m$ is integrated out:
$$
  {\cal L}_{YM} = -{1 \over 4\k^2}F_{\m \n}^2.
 \eqn\YML
$$

If $\k = 0$, the Lagrangian \START\ would have gauge symmetry
$$
  \da A_\m = 0, \quad \da B_\m = D_\m C
 \eqn\BFG
$$
in addition to the nonabelian gauge invariance
$$
  \db A_\m = D_\m c,
   \quad \db B_\m = -ig[B_\m, c],
 \eqn\NAG
$$
where $D_\m = \p_\m - ig[A_\m, \ \ ]$ and
the gauge-transformation parameters $C$ and $c$ take values in
the algebra $\cal G$.

In order to retain the additional gauge symmetry \BFG\ also for
$\k \neq 0$, we introduce a St{\" u}ckelberg-like
field $\L$,
taking values in $\cal G$,
which transforms as
$$
  \da \L = \k C, \quad \db \L = -ig[\L, c],
 \eqn\LAMBDA
$$
and rewrite $B_\m$ into $B_\m - \k^{-1} D_\m \L$
in the Lagrangian \START\ to obtain
$$
  {\cal L}_C = {1 \over 2}\e^{\mu \nu \r}B_\m F_{\n \r}
                + {1 \over 2}\k^2(B_\m - \k^{-1} D_\m \L)^2
 \eqn\NEXT
$$
with the help of the Bianchi identity
to the curvature $F_{\m \n}$.
This Lagrangian is gauge equivalent to the Lagrangian \START\
by construction.

%
{\caps 3. Covariant Gauge-Fixing}

Now that we have gotten the gauge symmetry $\d = \da + \db$
for the Lagrangian \NEXT\ in the previous section,
we can proceed to construct a gauge-fixed Lagrangian
by means of the BRS transformation $\brs$ corresponding to $\d$.

We first regard the parameters $C$ and $c$ as fermionic FP ghosts
whose transformation law is given by
$$
  \da C = 0, \quad \db C = ig[C, c];
  \quad \da c = 0, \quad \db c = {i \over 2}g[c, c]
 \eqn\FPT
$$
so as to satisfy the nilpotency of $\brs = \da + \db$.

We further introduce FP anti-ghosts ${\bar C}$, ${\bar c}$ and
NL fields $B$, $b$ that take values in the Lie algebra $\cal G$
and obey the transformation rule
$$
 \eqalign{
  &\da {\bar C} = iB, \quad \db {\bar C} = ig[{\bar C}, c];
   \quad \da B = 0, \quad \db B = -ig[B, c]; \cr
  &\da {\bar c} = 0, \quad \db {\bar c} = ib;
   \quad \da b = 0, \quad \db b = 0, \cr
 }
 \eqn\BRS
$$
which also keeps the nilpotency of $\brs$ intact.
More precisely, the two transformations $\da$ and $\db$ independently
satisfy the nilpotency.

Then we obtain a gauge-fixed Lagrangian
$$
 \eqalign{
  {\cal L} &= {\cal L}_C - i\brs ({\bar C}D^\m B_\m
               + {\bar c}\{ \p^\m A_\m + {1 \over 2}\a b \}) \cr
           &= {\cal L}_C - i\da ({\bar C}D^\m B_\m)
               -i\db ({\bar c}\{ \p^\m A_\m + {1 \over 2}\a b \}), \cr
 }
 \eqn\GFL
$$
where $\a$ denotes a gauge parameter.
This Lagrangian takes the form
$$
 \eqalign{
  {\cal L} &= {1 \over 2}\e^{\mu \nu \r}B_\m F_{\n \r}
               + {1 \over 2}(D_\m \L)^2
                + {1 \over 2}\k^2 B_\m^2 \cr
           &\quad + ND^\m B_\m +i{\bar C}D^\m D_\m C
            + b\p^\m A_\m +{1 \over 2}\a b^2
             + i{\bar c}\p^\m D_\m c, \cr
 }
 \eqn\LAGL
$$
up to total derivative, in terms of the redefined field
$N = B + \k \L$
with its BRS transformation law
$$
  \brs {\bar C} = i(N - \k \L + g[{\bar C}, c]),
   \quad \brs N = \k^2 C -ig[N, c].
 \eqn\MBRS
$$

\endpage

%
{\caps 4. Perturbative Expansion}

We are ready to consider a new perturbation scheme
in the nonabelian gauge theory \LAGL.
Our proposal is to treat the term ${1 \over 2}\k^2 B_\m^2$
as an interaction one rather than a kinetic one though it is quadratic
in the fields.
Hence the kinetic terms are given by
$$
 \eqalign{
  {\cal L}_K &= \e^{\mu \nu \r}B_\m \p_\n A_\r
                 + {1 \over 2}(\p_\m \L)^2 \cr
             &\quad + N\p^\m B_\m +i{\bar C}\p^\m \p_\m C
              + b\p^\m A_\m +{1 \over 2}\a b^2
               + i{\bar c}\p^\m \p_\m c, \cr
 }
 \eqn\KINET
$$
and the interaction ones by ${\cal L} - {\cal L}_K$.
The propagators of the vector fields $A_\m$ and $B_\n$
are obtained as
$$
  {1 \over k^2}\pmatrix{\a {k_\m k_{\m '} \over k^2}
           & -i\e_{\m {\n '} \r}k^\r \cr
           -i\e_{\n {\m '} \r}k^\r & 0 \cr}.
 \eqn\PROP
$$

In the (ultraviolet) region where both of the two couplings
$g$ and $\k$ are weak, we can deal with the full interaction
${\cal L} - {\cal L}_K$ as perturbation.
Then the perturbative expansion based on the free theory \KINET\
essentially reproduces the same results for correlators
of the fields $A_\m$, $b$, $c$, and $\bar c$
as the ordinary perturbation theory provides.
We note that we need
$$
 \eqalign{
  \D {\cal L} &= -i\brs ({\bar C}{1 \over 2}\z \{ N + \k \L \}) \cr
              &= \z ({1 \over 2}N^2 - {1 \over 2}\k^2 \L^2
                  + i\k^2{\bar C}C) \cr
 }
 \eqn\COUNT
$$
as a counter term to be added to the Lagrangian \LAGL,
where $\z$ denotes another gauge parameter.

Let us turn to the consideration of the infrared regime.
In this section, we restrict ourselves to the case $\a \neq 0$,
leaving the investigation of the other
case $\a = 0$ to the next section.
The form of the propagators \PROP\ shows that the mass dimensions
of the fields $A_\m$ and $B_\m$ are one half and three halves,
respectively.
Thus the coupling $g^2$ has the dimension of mass, and the other
one $\k^2$ is dimensionless.

Since the renormalization constants for the theory \LAGL\
are finite except the one for the part \COUNT,
the coupling constants $g^2$ and $\k^2$ themselves do not run at all.
In particular, the dimensionless coupling $\k^2$ can continue to be
small even in the infrared regime.
However, the coupling $g^2$ behaves as $e^{-t}g^2$
due to its mass dimension
when the relevant momentum $p_\m$ is scaled as $e^tp_\m$.
Hence, in the infrared regime, the interaction terms involving
the coupling $g$ should be treated nonperturbatively,
while the term ${1 \over 2}\k^2 B_\m^2$ may be regarded
as small perturbation;
accordingly nonperturbative information of YM theory
might be partially incorporated in that of the BF theory
coupled to the St{\" u}ckelberg-like field
$$
 \eqalign{
  {\cal L}_{BF} &= {1 \over 2}\e^{\mu \nu \r}B_\m F_{\n \r}
                    + {1 \over 2}(D_\m \L)^2 \cr
                &\quad + ND^\m B_\m +i{\bar C}D^\m D_\m C
                 + b\p^\m A_\m +{1 \over 2}\a b^2
                  + i{\bar c}\p^\m D_\m c. \cr
 }
 \eqn\BFL
$$
We also suspect that formidable infrared divergences present in
the conventional perturbative expansion in
YM theory might be cured considerably in the perturbation scheme
based on nonperturbative treatment of the zeroth-order theory \BFL.
(In this connection, see discussion in the final section.)

%
{\caps 5. Landau-Gauge Peculiarity}

In the previous section, we have concentrated on the case $\a \neq 0$
for the perturbative expansion in the infrared regime.
The reason why the case $\a = 0$ should be considered
separately is that the Landau gauge has peculiarity
which invalidates the lines of reasoning
that led to weakness of the coupling $\k$ for the case $\a \neq 0$.

The Landau-gauge
peculiarity manifests itself in the form of the propagators \PROP.
When $\a = 0$, only the transition propagators between $A_\m$ and
$B_\n$ are non-vanishing.
This makes it meaningless to consider the separate normalizations
of the two fields $A_\m$ and $B_\n$, and thus
the independent sizes of the two couplings $g$ and $\k$.

Indeed the model \BFL\ is one-loop exact in the usual perturbation
theory\rlap.
\refmark{\Oda}
Then the perturbative treatment of the term ${1 \over 2}\k^2 B_\m^2$
based on the BF theory
essentially reproduces the same result as
what is given by the ordinary perturbation theory,
which is applicable only in the ultraviolet regime.
In the infrared regime,
we need nonperturbative treatment of the interaction
${1 \over 2}\k^2 B_\m^2$ and hence the full YM theory
in the Landau gauge.

\endpage

%
{\caps 6. Discussion}

Turning back to the case $\a \neq 0$,
we are led to ask whether nonperturbative contents
of the BF theory \BFL\
bear any essential features of the original YM theory.
Fortunately enough, we have remarkable evidence
that strongly suggests the affirmative answer.

Let us integrate out the fields ${\bar C}$, $C$, $N$, and $B_\m$
sequentially in the Lagrangian \BFL\ to get
$$
  {\cal L}_T = {1 \over 2}(D_\m \L)^2
                + b\p^\m A_\m +{1 \over 2}\a b^2
                 + i{\bar c}\p^\m D_\m c
 \eqn\PGM
$$
with the constraint $F_{\m \n} = 0$,
which can be solved as $gA_\m = -i(\p_\m {\sl g}){\sl g}^{-1}$
by means of a field ${\sl g}$ taking values in the gauge group
corresponding to the algebra $\cal G$.
This reveals that the theory \PGM, or \BFL, is essentially
a three-dimensional analogue of Hata's pure-gauge model
\refmark{\Hat}
in four dimensions.

A slightly modified version of the pure-gauge model
has been shown
\refmark{\Kug}
to satisfy Kugo-Ojima's sufficient condition
\refmark{\Oji}
for color confinement in nonabelian gauge theory.
This is of nonperturbative origin
with nonlinearity of the group-valued field ${\sl g}$
playing an important role to realize that.
Note that the Landau-gauge peculiarity can also be understood
from the present viewpoint since the Landau gauge suppresses
the crucial pure-gauge fluctuations completely.
We further remark that
importance of the pure-gauge part in covariant formalism
of YM theory is clear from its contribution to the asymptotic freedom
in the regime of deep inelastic scattering.

The above considerations imply that the perturbation scheme
proposed in this paper might be consistent with the confining nature
of nonabelian gauge theory.
We of course need more investigation on this aspect
in four spacetime dimensions.

%
\acknowledge

The author would like to thank M.~Abe, H.~Hata, N.~Ikeda,
and S.~Yahikozawa
for fruitful discussions. He is also grateful to T.~Kugo for
helpful discussions and
careful reading of the manuscript.

\refout

\bye